\documentclass{iopart}

\usepackage{braket}
\renewcommand{\vec}[1]{\mathbf{#1}}

\begin{document}
\bibliographystyle{iopart-num}
\title[\jpa]{Quantum Field Theory of Electrons and Nuclei}

\author{Ville J. H\"{a}rk\"{o}nen}

\address{Computational Physics Laboratory, Tampere University, P.O. Box 692, FI-33014 Tampere, Finland}
\ead{ville.j.harkonen@gmail.com}


\vspace{10pt}
\begin{indented}
\item[]March 2024
\end{indented}

\begin{abstract}
We develop a non-relativistic quantum field theory of electrons and nuclei based on the Coulomb Hamiltonian. We derive the exact equations of motion and write these equations in the form of Hedin's equations for all species of identical particles involved. Theory derived allows the computation of exact observables and provides a rigorous starting point to derive approximations in a systematic way.
\end{abstract}

%
\vspace{2pc}
\submitto{\jpa}
%
%
%

\section{Introduction}
\label{Introduction}

The description of molecules and solids, comprising electrons and nuclei, is commonly based on the Coulomb Hamiltonian. Moreover, the description of these systems heavily rely on the Born-Oppenheimer approximation \cite{Born-Oppenheimer-Adiabatic-Approx.1927,born-huang-dynamical-1954}, but also various approaches beyond it have been developed \cite{Kreibich-MulticompDFTForElectronsAndNuclei-PhysRevLett.86.2984-2001,Kreibich-MulticompDFTForElectronsAndNuclei-PhysRevA.78.022501-2008,Gidopoulos-Gross-ElectronicNonAdiabaticStates-2014,Abedi-ExactFactorization-PhysRevLett.105.123002-2010,Baym-field-1961,Harkonen-ManybodyGreensFunctionTheoryOfElectronsAndNucleiBeyondTheBornOppenheimerApproximation-PhysRevB.101.235153-2020,Harkonen-ExactFactorizationOfTheManyBodyGreensFunctionTheoryOfElectronsAndNuclei-PhysRevB.106.205137-2022}. The field theoretical many-body Green's function approach \cite{Baym-field-1961,Harkonen-ManybodyGreensFunctionTheoryOfElectronsAndNucleiBeyondTheBornOppenheimerApproximation-PhysRevB.101.235153-2020,Harkonen-ExactFactorizationOfTheManyBodyGreensFunctionTheoryOfElectronsAndNuclei-PhysRevB.106.205137-2022} treats the electrons and nuclei differently. Namely, the electronic operators are written in terms of field operators and the nuclear variables are in first quantization. Treating the nuclei in first quantization has been very successful approach in explaining the properties of electron-nuclear systems like molecules and solids. The nuclear problem can be exactly solved within the harmonic approximation when the nuclear coordinates in the nuclear Hamiltonian are expanded up to second order about the equilibrium positions. The resulting quadratic Hamiltonian can be diagonalized with various methods \cite{born-huang-dynamical-1954,Harkonen-OnTheDiagonalizationOfQuadraticHamiltonians-2021}. This approach is well established for many situations in which the nuclei are rather localized close to their equilibrium positions and consequently there are no significant overlap between nuclear densities. The theory of molecular vibrations and lattice dynamics is well-developed also from a computational point of view and the open source computational packages like Quantum Espresso \cite{Baroni-PhononsAndRelatedCrystalPropFromDFTPT-RevModPhys.73.515-2001,Giannozzi-QuantumEspresso-2009} can be used to compute all the quantities needed to solve the nuclear problem. By doing so, we can compute a number of nuclear observables, many of them matching well to the experimental values \cite{Harkonen-NTE-2014,Harkonen-Tcond-II-VIII-PhysRevB.93.024307-2016,Harkonen-AbInitioComputStudyOnTheLattThermalCondOfZintlClathrates-PhysRevB.94.054310-2016,Tadano-FirstPrinciplesPhononQuasiparticleTheoryAppliedToAStronglyAnharmonicHalidePerovskite-PhysRevLett.129.185901-2022}.

Despite the success of the first quantization approach for the nuclei, we may still ask, what is the non-relativistic quantum field theory \cite{Baiguera-AspectsOfNonRelativisticQuantumFieldTheories-2024} of electron-nuclei many-body systems such that all degrees of freedom are described by field operators. Such a theory could turn out convenient in some situations, particularly in those when the nuclei are not well localized \cite{Benoit-TunnellingAndZeroPointMotionInHighPressureIce-1998,Benton-ClassicalAndQuantumTheoriesOfProtonDisorderInHexagonalWaterIce-PhysRevB.93.125143-2016}. Further, in systems where the Born-Oppenheimer approximation breaks down \cite{Vidal-EvidenceOnTheBreakdownOfTheBornOppenheimerApproximationInTheChargeDensityOfCrystalline7LiHD-1992,Harkonen-BreakdownOfTheBornOppenheimerApproximationInSolidHydrogenAndHydrogenRichSolids-Arxiv-2023,Harkonen-BreakdownOfTheBornOppenheimerApproximationInLiHandLiD-Arxiv-2023} the coupling of the electron and nuclear equations become important \cite{Harkonen-ManybodyGreensFunctionTheoryOfElectronsAndNucleiBeyondTheBornOppenheimerApproximation-PhysRevB.101.235153-2020} and our approach developed here provides an alternative way to deal with this coupling. In this work we derive the exact non-relativistic quantum field theory of electrons and nuclei given the Coulomb Hamiltonian. We show that the resulting equations can be written in the form of the Hedin's equations \cite{Hedin-NewMethForCalcTheOneParticleGreensFunctWithApplToTheElectronGasProb-PhysRev.139.A796-1965} for all different species of identical particles (electrons and different species of nuclei).

This work is organized as follows. We write the Hamiltonian in terms of field operators in Sec. \ref{Hamiltonian}. We define the Green's functions in Sec. \ref{GreensFunctions}. The exact equations of motion for the Green's functions and related quantities are considered in Sec. \ref{EquationsOfMotion}.

\section{Hamiltonian}
\label{Hamiltonian}

We consider a system of $N_{e}$ electrons and $N_{n}$ nuclei. The position coordinates of these particles $\vec{r}_{i}, \ i = 1,\ldots,N_{e},$ and $\vec{R}_{j}, \ j = 1,\ldots,N_{n}$. We assume that these particle interactions are Coulombic and thus the starting point is the Hamiltonian $H = T + V$, where $T$ is the kinetic energies of the electrons and nuclei and $V$ the potential energy originating from the Coulomb force. More explicitly, the Hamiltonian can be written as
\begin{equation} 
H = T_{n} + T_{e} + V_{ee} + V_{en} + V_{nn},
\label{eq:HamiltonianEq_1}
\end{equation}
where the kinetic energies in position representation are
\begin{equation} 
T_{n}  = - \sum^{N_{n}}_{j = 1} \frac{ \hbar^{2} }{ 2 M_{j} } \nabla^{2}_{\vec{R}_{j}}, \quad T_{e} = - \frac{ \hbar^{2} }{ 2 m_{e} } \sum^{N_{e}}_{i = 1}  \nabla^{2}_{\vec{r}_{i}}.
\label{eq:HamiltonianEq_2}
\end{equation}
The potential energy contributions can be written as
\begin{eqnarray} 
V_{ee}  &=& \frac{1}{2} \sum^{N_{e}}_{ i,i'= 1 }{}^{'} \frac{ \varsigma }{ \left| \vec{r}_{i} - \vec{r}_{i'} \right| }, \quad V_{en} = \sum^{N_{e}}_{ i= 1 }\sum^{N_{n}}_{ j= 1 } \frac{ - Z_{j} \varsigma }{ \left| \vec{r}_{i} - \vec{R}_{j} \right| }, \nonumber \\
V_{nn}  &=& \frac{1}{2} \sum^{N_{n}}_{ j,j'= 1 }{}^{'} \frac{ Z_{j} Z_{j'} \varsigma }{ \left| \vec{R}_{j} - \vec{R}_{j'} \right| }.
\label{eq:HamiltonianEq_3}
\end{eqnarray}
Here the primed sums are over the values $i \neq i'$, $j \neq j'$ and $\varsigma \equiv e^{2}/ \left( 4 \pi \epsilon_{0} \right)$. The Hamiltonian of Eq. \ref{eq:HamiltonianEq_1} is invariant under the translations and rotations of all particles. This leads to subtle issues as discussed in earlier works \cite{Sutcliffe-TheDecouplingOfElectronicAndNuclearMotions-2000,Kreibich-MulticompDFTForElectronsAndNuclei-PhysRevLett.86.2984-2001,Harkonen-ManybodyGreensFunctionTheoryOfElectronsAndNucleiBeyondTheBornOppenheimerApproximation-PhysRevB.101.235153-2020}. These issues, however, can vanish when simplifications, like the Born-Oppenheimer approximation \cite{born-huang-dynamical-1954} are established. These mentioned issues can be solved by changing the frame of reference from laboratory frame (in which the equations above are written) to a body-fixed frame \cite{Sutcliffe-TheDecouplingOfElectronicAndNuclearMotions-2000,Kreibich-MulticompDFTForElectronsAndNuclei-PhysRevLett.86.2984-2001,Harkonen-ManybodyGreensFunctionTheoryOfElectronsAndNucleiBeyondTheBornOppenheimerApproximation-PhysRevB.101.235153-2020}. To concentrate on the new approach and to keep the notation as simple as possible, we retain to laboratory frame formulation while we acknowledge its limitations. The Coulomb part of the problem in body-fixed frame is still formally of the same form \cite{Harkonen-ManybodyGreensFunctionTheoryOfElectronsAndNucleiBeyondTheBornOppenheimerApproximation-PhysRevB.101.235153-2020} as it is in the laboratory frame. Thus, the results of this work can be transformed in many cases to those in the body-fixed frame.

To setup a many-body Green's function theory of electrons and nuclei, the following approach is conventionally taken \cite{Baym-field-1961,Harkonen-ManybodyGreensFunctionTheoryOfElectronsAndNucleiBeyondTheBornOppenheimerApproximation-PhysRevB.101.235153-2020,Harkonen-ExactFactorizationOfTheManyBodyGreensFunctionTheoryOfElectronsAndNuclei-PhysRevB.106.205137-2022}: write the electronic variables in terms of field operators, treat the nuclei in first quantization. After that the equations of motion are written for the electronic field operators, nuclear displacement operators and the corresponding Green's functions, from which the observables can be extracted. Here we take a different route and treat electrons and different species of nuclei, on the same footing and describe all species in terms of field operators. By using the usual procedure of field theory \cite{Fetter-Walencka-q-theory-of-many-particle-1971,Mahan-many-particle-1990,Gross-ManyParticleTheory-1991} to all species of identical particles involved, the Hamiltonian in second quantized form can then be written as
\begin{eqnarray} 
\hat{H}\left(t\right) &=& \sum^{N_{s}}_{k = 1} \int d\vec{r} \hat{\psi}^{\dagger}_{k}\left(\vec{r},t\right) D_{k}\left(\vec{r}t\right) \hat{\psi}_{k}\left(\vec{r},t\right) \nonumber \\
&&+ \frac{1}{2} \sum^{N_{s}}_{k = 1} \int d\vec{r} \int d\vec{r}'  v_{kk}\left(\vec{r},\vec{r}'\right) \hat{\psi}^{\dagger}_{k}\left(\vec{r}\right) \hat{\psi}^{\dagger}_{k}\left(\vec{r}'\right) \hat{\psi}_{k}\left(\vec{r}'\right) \hat{\psi}_{k}\left(\vec{r}\right) \nonumber \\
&&+ \frac{1}{2} \sum^{N_{s}}_{k , k' = 1}{}^{'} \int d\vec{r} \int d\vec{r}'  v_{kk'}\left(\vec{r},\vec{r}'\right) \hat{n}_{k}\left(\vec{r}\right) \hat{n}_{k'}\left(\vec{r}'\right),
\label{eq:HamiltonianEq_5}
\end{eqnarray}
where $N_{s}$ is the number of species and
\begin{eqnarray}
\hat{n}_{k}\left(\vec{r}\right) &\equiv& \hat{\psi}^{\dagger}_{k}\left(\vec{r}\right) \hat{\psi}_{k}\left(\vec{r}\right), \nonumber \\
D_{k}\left(\vec{r}t\right) &\equiv& -\frac{\hbar^{2} }{2 m_{k}} \nabla^{2}_{k} + \varphi\left(\vec{r}t\right) Z_{k} + F_{k}\left(\vec{r}t\right), \nonumber \\
v_{kk'}\left(\vec{r},\vec{r}'\right) &\equiv& Z_{k} Z_{k'} \chi_{kk'} v\left(  \vec{r}, \vec{r}' \right), \quad v\left(  \vec{r}, \vec{r}' \right) = \varsigma / \left| \vec{r} - \vec{r}' \right|.
\label{eq:HamiltonianEq_6}
\end{eqnarray}
Here, the indices $k,k'$ appear as subscripts in the interaction potentials in order to assign appropriate charges for different species and Here $Z_{k}$ is the electric charge of the species $k$. Moreover, $\chi_{kk'}$ in Eq. \ref{eq:HamiltonianEq_6} has the value $1$ for $k=k'$ and the value $2$ for $k \neq k'$. We have added the external time-dependent potentials $\varphi\left(\vec{r}t\right) Z_{k}$ and $F_{k}\left(\vec{r}t\right)$ to use the functional derivative approach \cite{Schwinger-OnTheGreensFunctionsOfQuantizedFieldsI-1951} in deriving the equations of motion. These potentials are set to zero at the end so that the Hamiltonian of Eq. \ref{eq:HamiltonianEq_5} gives the matrix elements $H$ given by Eq. \ref{eq:HamiltonianEq_1}. The field operators in Eq. \ref{eq:HamiltonianEq_5} satisfy the following (anti)commutation relations
\begin{equation} 
\left[\hat{\psi}_{k}\left(\vec{r},t\right),\hat{\psi}^{\dagger}_{k}\left(\vec{r}',t\right)\right]_{\pm} =  \delta\left(\vec{r}-\vec{r}'\right),
\label{eq:HamiltonianEq_7}
\end{equation}
and
\begin{equation} 
\left[\hat{\psi}^{\dagger}_{k}\left(\vec{r},t\right),\hat{\psi}^{\dagger}_{k}\left(\vec{r}',t\right)\right]_{\pm} = \left[\hat{\psi}_{k}\left(\vec{r},t\right),\hat{\psi}_{k}\left(\vec{r}',t\right)\right]_{\pm} = 0.
\label{eq:HamiltonianEq_8}
\end{equation}
That is, for fermionic particles the field operators anti-commute and for bosons the commutation relations are satisfied. If the particles are of a different species, the (anti)commutation relations are the following. Let $k$ denote bosonic species and $k'$ fermionic species or let $k$ and $k'$ denote two different bosonic species or two different fermionic species (that is, in all cases when $k \neq k'$), then
\begin{eqnarray} 
\left[\hat{\psi}_{k}\left(\vec{r},t\right),\hat{\psi}^{\dagger}_{k'}\left(\vec{r}',t\right)\right]_{-} &=& \left[\hat{\psi}_{k'}\left(\vec{r},t\right),\hat{\psi}^{\dagger}_{k}\left(\vec{r}',t\right)\right]_{-} = 0, \nonumber \\
\left[\hat{\psi}^{\dagger}_{k}\left(\vec{r},t\right),\hat{\psi}^{\dagger}_{k'}\left(\vec{r}',t\right)\right]_{-} &=& \left[\hat{\psi}_{k}\left(\vec{r},t\right),\hat{\psi}_{k'}\left(\vec{r}',t\right)\right]_{-} = 0.
\label{eq:HamiltonianEq_9}
\end{eqnarray}
We have now all necessary results to define the Green's functions and to derive the equations of motion for them.

\section{Green's Functions}
\label{GreensFunctions}

The one-body Green's function for a species $k$ is defined as
\begin{equation} 
G_{k}\left(\vec{r}t,\vec{r}'t'\right) \equiv - \frac{i}{\hbar}\left\langle  \mathcal{T} \left\{\hat{\psi}_{k}\left(\vec{r}t\right) \hat{\psi}^{\dagger}_{k}\left(\vec{r}'t'\right) \right\} \right\rangle,
\label{eq:GreensFunctionsEq_1}
\end{equation}
where
\begin{eqnarray} 
\left\langle  \mathcal{T} \left\{\hat{\psi}_{k}\left(\vec{r}t\right) \hat{\psi}^{\dagger}_{k}\left(\vec{r}'t'\right) \right\} \right\rangle &=& \theta\left(t-t'\right) \left\langle  \hat{\psi}_{k}\left(\vec{r}t\right) \hat{\psi}^{\dagger}_{k}\left(\vec{r}'t'\right) \right\rangle \nonumber \\
&&- \theta\left(t'-t\right) \left\langle \hat{\psi}^{\dagger}_{k}\left(\vec{r}'t'\right) \hat{\psi}_{k}\left(\vec{r}t\right)  \right\rangle.
\label{eq:GreensFunctionsEq_2}
\end{eqnarray}
The field operators $\hat{\psi}_{k}\left(\vec{r}t\right), \hat{\psi}^{\dagger}_{k}\left(\vec{r}'t'\right)$ are assumed to be operators in the Heisenberg picture and the subscript $H$ is neglected for the sake of notational convenience. The ensemble averages in Eqs. \ref{eq:GreensFunctionsEq_1} and \ref{eq:GreensFunctionsEq_2} are of the form
\begin{equation} 
\left\langle \hat{o}\left(t\right) \right\rangle = \sum_{n} \braket{\Psi_{n}|\hat{\rho} \hat{o}\left(t\right) |\Psi_{n}} = Tr\left[\hat{\rho} \hat{o}\left(t\right) \right],
\label{eq:GreensFunctionsEq_3}
\end{equation}
and the density operator is the grand canonical statistical operator
\begin{equation} 
\hat{\rho} =  \frac{e^{-\beta \hat{H}^{M} }}{Z}, \quad Z= Tr\left[e^{-\beta \hat{H}^{M} }\right],
\label{eq:GreensFunctionsEq_4}
\end{equation}
where
\begin{equation} 
\hat{H}^{M} = \hat{H} - \mu_{e} \hat{N}_{e}, \quad \hat{N}_{e} = \int d\vec{r} \hat{n}_{1}\left(\vec{r}\right),
\label{eq:GreensFunctionsEq_5}
\end{equation}
and $\mu_{e}$ is the chemical potential of the electrons. We thus assumed that the electrons are species $k = 1$.

\section{Equations of Motion}
\label{EquationsOfMotion}

We start by writing the Heisenberg equations of motion for the field operator
\begin{equation} 
i \hbar \frac{\partial}{\partial{t}} \hat{\psi}_{k}\left(\vec{r},t\right) = \left[\hat{\psi}_{k}\left(\vec{r},t\right),\hat{H}\left(t\right)\right]_{-}.
\label{eq:EquationsOfMotionEq_1}
\end{equation}
After computing the commutator of Eq. \ref{eq:EquationsOfMotionEq_1} we write
\begin{eqnarray} 
i \hbar \frac{\partial}{\partial{t}} \hat{\psi}_{k}\left(\vec{r}t\right) &=& \sum^{N_{s}}_{k'= 1}{}^{'} \int d\vec{r}'  v_{kk'}\left(\vec{r},\vec{r}'\right) \hat{\psi}_{k}\left(\vec{r}t\right) \hat{n}_{k'}\left(\vec{r}'t\right) \nonumber \\
&&+ \int d\vec{r}'  v_{kk}\left(\vec{r},\vec{r}'\right) \hat{\psi}^{\dagger}_{k}\left(\vec{r}'t\right) \hat{\psi}_{k}\left(\vec{r}'t\right) \hat{\psi}_{k}\left(\vec{r}t\right) \nonumber \\
&&+D_{k}\left(\vec{r}t\right) \hat{\psi}_{k}\left(\vec{r}t\right).
\label{eq:EquationsOfMotionEq_2}
\end{eqnarray}
By using Eq. \ref{eq:EquationsOfMotionEq_2} we can start writing the equations of motion for the Green's functions of Eq. \ref{eq:GreensFunctionsEq_1} and determining equations for the related quantities. We derive these equations in \ref{AuxillaryResults} and we have obtained a set of equations for the system of electrons and nuclei with Coulomb interactions with no further approximations. To summarize the results (Eqs. \ref{eq:EquationsOfMotionEq_26}, \ref{eq:EquationsOfMotionEq_25}, \ref{eq:QuantumFieldTheoryInteractSchrodingerFieldForSevDiffKindOfIdPartSIndGreensFunctVertexFunctEq_13}, \ref{eq:QuantumFieldTheoryInteractSchrodingerFieldForSevDiffKindOfIdPartSIndGreensFunctScrCoulIntEq_3}, \ref{eq:QuantumFieldTheoryInteractSchrodingerFieldForSevDiffKindOfIdPartSIndGreensFunctPolarizEq_6}, respectively)
\begin{eqnarray} 
\delta\left(1 - 2\right) &=& \left[i \hbar\frac{\partial}{\partial{t_{1}}} + \frac{\hbar^{2} }{2 m_{k}} \nabla^{2}_{k} - V_{tot}\left(1,k\right) \right] G_{k}\left(1,2\right), \nonumber \\
&&- \int d3  \Sigma_{k}\left(1,3\right) G_{k}\left(3,2\right), \nonumber \\
\Sigma_{k}\left(1,4\right)  &=& i \hbar \int d3 \int d5 W_{k}\left(1,5\right)  G_{k}\left(1,3\right)  \Gamma_{k}\left(3,4,5\right), \nonumber \\
\Gamma_{k}\left(1,2,3\right)&=& \delta\left(1 - 2\right) \delta\left(1 - 3\right) + \int d4\int d5 \int d6 \int d7 \nonumber \\
                            &&\ \ \ \times \frac{\delta{\Sigma_{k}\left(1,2\right)}}{\delta{G_{k}\left(4,5\right)}} G_{k}\left(4,6\right) G_{k}\left(7,5\right) \Gamma_{k}\left(6,7,3\right), \nonumber \\
 			W_{k}\left(1,2\right) &=& Z^{2}_{k} v\left(1,2\right) \nonumber \\
			&&+  Z^{2}_{k} \int d3 \int d4  v\left(2,3\right) \sum_{k'} P_{k'}\left(3,4\right)  W_{k'}\left(1,4\right), \nonumber \\
P_{k}\left(1,2\right)       &=& -i \hbar \int d3 \int d4  G_{k}\left(1,3\right) G_{k}\left(4,1^{+}\right) \Gamma_{k}\left(3,4,2\right). 
\label{eq:HedinLikeEquationsEq_1}
\end{eqnarray}
We call the set of equations given in Eq. \ref{eq:HedinLikeEquationsEq_1} the multispecies Hedin's equations. These equations are indeed formally the same as the ones for electrons within the Born-Oppenheimer approximation \cite{Hedin-NewMethForCalcTheOneParticleGreensFunctWithApplToTheElectronGasProb-PhysRev.139.A796-1965} and beyond it \cite{Harkonen-ManybodyGreensFunctionTheoryOfElectronsAndNucleiBeyondTheBornOppenheimerApproximation-PhysRevB.101.235153-2020}. The coupling of different species is through all terms of Eq. \ref{eq:HedinLikeEquationsEq_1}: the screened Coulomb interaction $W_{k}\left(1,2\right)$, the self-energy $\Sigma_{k}\left(1,4\right)$, the vertex function $\Gamma_{k}\left(1,2,3\right)$ and the polarization $P_{k}\left(1,2\right)$. We note that we have not imposed any approximations and the relations in Eq. \ref{eq:HedinLikeEquationsEq_1} are exact given the Coulomb Hamiltonian. The Green's functions are defined with respect to states of the full electron-nuclear space. The simplest case is when we have only one species of nuclei. In this case we only have two sets of equations $k = 1,2$: one set of equations \ref{eq:HedinLikeEquationsEq_1} for electrons and another for the identical nuclei. The same set of equations can thus be used to describe any system comprising the very same elements.

What is convenient from the practical point of view is that the Hedin's equations \cite{Hedin-NewMethForCalcTheOneParticleGreensFunctWithApplToTheElectronGasProb-PhysRev.139.A796-1965} have been a successful approach in deriving approximations to computationally solve the electronic problem within the Born-Oppenheimer approximation. In particular, the so-called GW-approximation \cite{Aryasetiawan-GeneralizedHedinsEquationsForQuantumManyBodySystemsWithSpinDependentInteractions-PhysRevLett.100.116402-2008} has become already a practical tool for solving various electronic properties of solids and molecules \cite{Nelson-SelfInteractInGreensFunctionTheoryOfTheHydrogenAtom-PhysRevA.75.032505-2007,Kutepov-ElectronicStructureSelfconsistentSolutionOfHedinsEquationsIncludingVertexCorrections-PhysRevB.94.155101-2016,Golze-TheGWCompendiumAPracticalGuideToTheoreticalPhotoemissionSpectroscopy-2019,Mejuto-Zaera--SelfConsistencyInGWGAMMAFormalismLeadingToQuasiparticleQuasiparticleCouplings-PhysRevB.106.165129-2022}. We can use the very same machinery to derive approximations to the multispecies equations. For example, the multispecies GW-approximation can be obtained by approximating $\Gamma_{k}\left(1,2,3\right) \approx \delta\left(1 - 2\right) \delta\left(1 - 3\right)$ for all $k$ and thus the approximate self-energy, screened Coulomb interaction and polarization follow from Eq. \ref{eq:HedinLikeEquationsEq_1}.

\section{Conclusions}
\label{Conclusions}

We have developed an exact non-relativistic quantum field theory of electrons and nuclei given the Coulomb Hamiltonian. We derived exact equations of motion which can be written in a form of the Hedin's equations. While the use of first quantization for the nuclei has been very useful for the description of molecules and solids in the past, we believe that our field theoretical formulation introduced here might be useful and convenient in some situations. These include systems where the nuclei are not well localized and we need to incorporate the (anti)symmetrization of states of identical particles to the formalism. Here, the (anti)symmetrization is automatically included through the use of field operators, also for the nuclei. The fact that an extensive amount of effort has been put into solving formally similar equations for electrons alone in the past five decades or so might render Eq. \ref{eq:HedinLikeEquationsEq_1} beneficial from a computational point of view. We also note that the breakdown of the Born-Oppenheimer approximation induces coupling between the electron and nuclear equations. Here we derived an alternative approach, distinct from the conventional methodology, to describe the electron-nuclei systems and to incorporate this coupling.

\appendix
\section{Auxillary Results}
\label{AuxillaryResults}

\subsection{Derivation of Equations of Motion}
\label{DerivationOfEquationsOfMotion}

By using Eq. \ref{eq:EquationsOfMotionEq_2}, the equations of motion for the Green's function of Eq. \ref{eq:GreensFunctionsEq_1} can be written as
\begin{eqnarray} 
&&\left[i \hbar\frac{\partial}{\partial{t}} - D_{k}\left(\vec{r}t\right) \right] G_{k}\left(\vec{r}t,\vec{r}'t'\right) =  \delta\left(t - t'\right) \delta\left(\vec{r} - \vec{r}'\right) \nonumber \\
&&- \frac{i}{ \hbar} \sum^{N_{s}}_{k'= 1} \int d\vec{r}''  v_{kk'}\left(\vec{r},\vec{r}''\right) \left\langle  \mathcal{T} \left\{ \hat{n}_{k'}\left(\vec{r}''t\right) \hat{\psi}_{k}\left(\vec{r}t\right) \hat{\psi}^{\dagger}_{k}\left(\vec{r}'t'\right) \right\} \right\rangle.
\label{eq:EquationsOfMotionEq_3}
\end{eqnarray}
Define
\begin{equation} 
\hat{n}\left(\vec{r}t\right) \equiv \sum^{N}_{k= 1} Z_{k} \hat{n}_{k}\left(\vec{r}t\right),
\label{eq:EquationsOfMotionEq_4}
\end{equation}
which is the total charge density and thus one may write Eq. \ref{eq:EquationsOfMotionEq_3} as
\begin{eqnarray} 
&&\left[i \hbar\frac{\partial}{\partial{t}} - D_{k}\left(\vec{r}t\right) \right] G_{k}\left(\vec{r}t,\vec{r}'t'\right) =  \delta\left(t - t'\right) \delta\left(\vec{r} - \vec{r}'\right) \nonumber \\
&&- \frac{i}{ \hbar} \int d^{3}r''  v\left(\vec{r},\vec{r}''\right) Z_{k} \left\langle  \mathcal{T} \left\{ \hat{n}\left(\vec{r}''t\right) \hat{\psi}_{k}\left(\vec{r}t\right) \hat{\psi}^{\dagger}_{k}\left(\vec{r}'t'\right) \right\} \right\rangle.
\label{eq:EquationsOfMotionEq_5}
\end{eqnarray}
We use
\begin{eqnarray} 
\left\langle \mathcal{T}\left\{ \hat{n}\left(\vec{r}''t\right) \hat{\psi}_{k}\left(\vec{r}t\right) \hat{\psi}^{\dagger}_{k}\left(\vec{r}'t'\right) \right\} \right\rangle &=& \left\langle \mathcal{T} \left\{ \hat{\psi}_{k}\left(\vec{r}t\right) \hat{\psi}^{\dagger}_{k}\left(\vec{r}'t'\right) \right\} \right\rangle \left\langle \hat{n}\left(\vec{r}''t\right) \right\rangle \nonumber \\
&&+ i\hbar \frac{\delta{\left\langle \mathcal{T} \left\{ \hat{\psi}_{k}\left(\vec{r}t\right) \hat{\psi}^{\dagger}_{k}\left(\vec{r}'t'\right) \right\} \right\rangle}}{\delta{\varphi\left(\vec{r}''t\right)}},
\label{eq:EquationsOfMotionEq_6}
\end{eqnarray}
and write Eq. \ref{eq:EquationsOfMotionEq_5} as
\begin{eqnarray} 
&&\left[i \hbar\frac{\partial}{\partial{t}} + \frac{\hbar^{2} }{2 m_{k}} \nabla^{2}_{k} - \varphi\left(\vec{r}t\right) Z_{k} - F_{k}\left(\vec{r}t\right) - \int d^{3}r''  v\left(\vec{r},\vec{r}''\right) Z_{k} \left\langle \hat{n}\left(\vec{r}''t\right) \right\rangle \right.  \nonumber \\
&&- \left. i\hbar \int d^{3}r''  v\left(\vec{r},\vec{r}''\right) Z_{k} \frac{\delta{}}{\delta{\varphi\left(\vec{r}''t\right)}} \right] G_{k}\left(\vec{r}t,\vec{r}'t'\right) =  \delta\left(t - t'\right) \delta\left(\vec{r} - \vec{r}'\right). \nonumber \\
\label{eq:EquationsOfMotionEq_10}
\end{eqnarray}
Define
\begin{equation} 
V_{tot}\left(\vec{r}t,k\right) \equiv \varphi\left(\vec{r}t\right) Z_{k} + F_{k}\left(\vec{r}t\right) + \int d^{3}r''  v\left(\vec{r},\vec{r}''\right) Z_{k} \left\langle \hat{n}\left(\vec{r}''t\right) \right\rangle,
\label{eq:EquationsOfMotionEq_11}
\end{equation}
and use the notation $\vec{r} t \equiv 1, \ \delta\left(1-2\right) \equiv \delta\left(t-t'\right)\delta\left(\vec{r} - \vec{r}'\right)$ and so on. Thus, we can write Eq. \ref{eq:EquationsOfMotionEq_10} as
\begin{eqnarray} 
\delta\left(1 - 2\right) &=& \left[i \hbar\frac{\partial}{\partial{t_{1}}} + \frac{\hbar^{2} }{2 m_{k}} \nabla^{2}_{k} - V_{tot}\left(1,k\right) \right. \nonumber \\
&&- \left. i\hbar \int d3  v\left(1,3\right) Z_{k} \frac{\delta{}}{\delta{\varphi\left(3\right)}} \right] G_{k}\left(1,2\right),
\label{eq:EquationsOfMotionEq_14}
\end{eqnarray}
where
\begin{eqnarray} 
v\left(1,3\right) &\equiv& \delta\left(t - t''\right) v\left(\vec{r},\vec{r}''\right), \quad  G_{k}\left(1,2\right) \equiv G_{k}\left(\vec{r}t,\vec{r}'t'\right), \nonumber \\
V_{tot}\left(1,k\right) &=& Z_{k} \varphi\left(1\right) + F_{k}\left(1\right)  + Z_{k} \int d3  v\left(1,3\right) \left\langle  \hat{n}\left(3\right) \right\rangle.
\label{eq:EquationsOfMotionEq_15}
\end{eqnarray}
The so-called dielectric function for a species $k$, $\epsilon_{k}\left(1,2\right)$, is defined through the following relation for its inverse
\begin{equation} 
\epsilon^{-1}_{k}\left(1,2\right) \equiv \frac{\delta{V_{tot}\left(1,k\right)}}{\delta{\varphi\left(2\right)}} = Z_{k} \delta\left(1-2\right) + Z_{k} \int d3  v\left(1,3\right) \frac{\delta{\left\langle  \hat{n}\left(3\right) \right\rangle}}{\delta{\varphi\left(2\right)}}.
\label{eq:EquationsOfMotionEq_17}
\end{equation}
For the functional derivative appearing in Eq. \ref{eq:EquationsOfMotionEq_14}
\begin{eqnarray} 
\frac{\delta{G_{k}\left(1,2\right)}}{\delta{\varphi\left(3\right)}} &=& - \int d4 \int d5 \int d6 G_{k}\left(1,4\right)  \frac{\delta{G^{-1}_{k}\left(4,5\right)}}{\delta{V_{tot}\left(6,k\right)}} \nonumber \\
&&\times \frac{\delta{V_{tot}\left(6,k\right)}}{\delta{\varphi\left(3\right)}} G_{k}\left(5,2\right).
\label{eq:EquationsOfMotionEq_18}
\end{eqnarray}
Define
\begin{equation} 
\Gamma_{k}\left(4,5,6\right) \equiv - \frac{\delta{G^{-1}_{k}\left(4,5\right)}}{\delta{V_{tot}\left(6,k\right)}}.
\label{eq:EquationsOfMotionEq_19}
\end{equation}
By using Eqs. \ref{eq:EquationsOfMotionEq_19} and the definition given by Eq. \ref{eq:EquationsOfMotionEq_17} in \ref{eq:EquationsOfMotionEq_18}, we obtain
\begin{eqnarray} 
\frac{\delta{G_{k}\left(1,2\right)}}{\delta{\varphi\left(3\right)}} &=& \int d4 \int d5 \int d6 G_{k}\left(1,4\right) \Gamma_{k}\left(4,5,6\right) \nonumber \\
&&\times \epsilon^{-1}_{k}\left(6,3\right) G_{k}\left(5,2\right).
\label{eq:EquationsOfMotionEq_20}
\end{eqnarray}
By using Eq. \ref{eq:EquationsOfMotionEq_20} in Eq. \ref{eq:EquationsOfMotionEq_14}
\begin{eqnarray} 
\delta\left(1 - 2\right) &=& \left[i \hbar\frac{\partial}{\partial{t_{1}}} + \frac{\hbar^{2} }{2 m_{k}} \nabla^{2}_{k} - V_{tot}\left(1,k\right) \right] G_{k}\left(1,2\right) \nonumber \\
&&- i\hbar Z_{k} \int d3 \int d4 \int d5 \int d6 v\left(1,3\right) G_{k}\left(1,4\right) \Gamma_{k}\left(4,5,6\right) \nonumber \\
&&\times \epsilon^{-1}_{k}\left(6,3\right) G_{k}\left(5,2\right).
\label{eq:EquationsOfMotionEq_21}
\end{eqnarray}
Let further
\begin{equation} 
W_{k}\left(1,6\right) \equiv Z_{k} \int d3 v\left(1,3\right) \epsilon^{-1}_{k}\left(6,3\right),
\label{eq:EquationsOfMotionEq_22}
\end{equation}
and thus Eq. \ref{eq:EquationsOfMotionEq_21} can be written as (some relabeling of the variables)
\begin{eqnarray} 
\delta\left(1 - 2\right) &=& \left[i \hbar\frac{\partial}{\partial{t_{1}}} + \frac{\hbar^{2} }{2 m_{k}} \nabla^{2}_{k} - V_{tot}\left(1,k\right) \right] G_{k}\left(1,2\right) \nonumber \\
&&- i\hbar \int d3 \int d4 \int d5 W_{k}\left(1,5\right) G_{k}\left(1,3\right) \nonumber \\
&&\times \Gamma_{k}\left(3,4,5\right) G_{k}\left(4,2\right).
\label{eq:EquationsOfMotionEq_24}
\end{eqnarray}
Define the self-energy as
\begin{equation} 
\Sigma_{k}\left(1,4\right) \equiv i \hbar \int d3 \int d5 W_{k}\left(1,5\right)  G_{k}\left(1,3\right)  \Gamma_{k}\left(3,4,5\right),
\label{eq:EquationsOfMotionEq_25}
\end{equation}
and Eq. \ref{eq:EquationsOfMotionEq_24} becomes 
\begin{eqnarray} 
\delta\left(1 - 2\right) &=& \left[i \hbar\frac{\partial}{\partial{t_{1}}} + \frac{\hbar^{2} }{2 m_{k}} \nabla^{2}_{k} - V_{tot}\left(1,k\right) \right] G_{k}\left(1,2\right) \nonumber \\
&&- \int d3  \Sigma_{k}\left(1,3\right) G_{k}\left(3,2\right).
\label{eq:EquationsOfMotionEq_26}
\end{eqnarray}
We have obtained the equation of motion for the Green's function $G_{k}\left(1,2\right)$ for each species $k$.

\subsection{Vertex Function}
\label{VertexFunction}

Consider Eq. \ref{eq:EquationsOfMotionEq_26}. This equation can be rearranged by defining
\begin{equation} 
h'_{k}\left(1\right) \equiv i \hbar\frac{\partial}{\partial{t_{1}}} + \frac{\hbar^{2} }{2 m_{k}} \nabla^{2}_{k} - V_{tot}\left(1,k\right),
\label{eq:QuantumFieldTheoryInteractSchrodingerFieldForSevDiffKindOfIdPartSIndGreensFunctVertexFunctEq_2}
\end{equation}
and then multiplying Eq. \ref{eq:EquationsOfMotionEq_26} from the right with $G^{-1}_{k}\left(2,\bar{3}\right)$ and integrating
\begin{eqnarray} 
 \int d2 \delta\left(1-2\right) G^{-1}_{k}\left(2,\bar{3}\right) &=& \int d2 h'_{k}\left(1\right) G_{k}\left(1,2\right) G^{-1}_{k}\left(2,\bar{3}\right) \nonumber \\
&&- \int d2\int d3  \Sigma_{k}\left(1,3\right) G_{k}\left(3,2\right) G^{-1}_{k}\left(2,\bar{3}\right), \nonumber \\
\label{eq:QuantumFieldTheoryInteractSchrodingerFieldForSevDiffKindOfIdPartSIndGreensFunctVertexFunctEq_3}
\end{eqnarray}
which becomes (some relabeling)
\begin{equation} 
G^{-1}_{k}\left(1,2\right) = h'_{k}\left(1\right)\delta\left(1 - 2\right) - \Sigma_{k}\left(1,2\right).
\label{eq:QuantumFieldTheoryInteractSchrodingerFieldForSevDiffKindOfIdPartSIndGreensFunctVertexFunctEq_5}
\end{equation}
By taking a functional derivate of Eq. \ref{eq:QuantumFieldTheoryInteractSchrodingerFieldForSevDiffKindOfIdPartSIndGreensFunctVertexFunctEq_5}
\begin{eqnarray} 
\frac{\delta{G^{-1}_{k}\left(1,2\right)}}{\delta{V_{tot}\left(3,k\right)}} &=& \delta\left(1 - 2\right) \frac{\delta{h'_{k}\left(1\right)}}{\delta{V_{tot}\left(3,k\right)}} - \frac{\delta{\Sigma_{k}\left(1,2\right)}}{\delta{V_{tot}\left(3,k\right)}} \nonumber \\
&=& \delta\left(1 - 2\right) \delta\left(1 - 3\right) - \frac{\delta{\Sigma_{k}\left(1,2\right)}}{\delta{V_{tot}\left(3,k\right)}},
\label{eq:QuantumFieldTheoryInteractSchrodingerFieldForSevDiffKindOfIdPartSIndGreensFunctVertexFunctEq_6}
\end{eqnarray}
where the left hand side, by Eq. \ref{eq:EquationsOfMotionEq_19}, is the vertex function times minus unity. One may write (Eq. \ref{eq:QuantumFieldTheoryInteractSchrodingerFieldForSevDiffKindOfIdPartSIndGreensFunctVertexFunctEq_2} is used)
\begin{equation} 
\Gamma_{k}\left(1,2,3\right) = \delta\left(1 - 2\right) \delta\left(1 - 3\right) + \frac{\delta{\Sigma_{k}\left(1,2\right)}}{\delta{V_{tot}\left(3,k\right)}}.
\label{eq:QuantumFieldTheoryInteractSchrodingerFieldForSevDiffKindOfIdPartSIndGreensFunctVertexFunctEq_7}
\end{equation}
Since $\Sigma_{k}\left(1,2\right)$ is a functional of the Green's functions, the chain rule may be used for the term including self-energy, namely
\begin{equation} 
\frac{\delta{\Sigma_{k}\left(1,2\right)}}{\delta{V_{tot}\left(3,k\right)}} = \int d4\int d5 \frac{\delta{\Sigma_{k}\left(1,2\right)}}{\delta{G_{k}\left(4,5\right)}} \frac{\delta{G_{k}\left(4,5\right)}}{\delta{V_{tot}\left(3,k\right)}}.
\label{eq:QuantumFieldTheoryInteractSchrodingerFieldForSevDiffKindOfIdPartSIndGreensFunctVertexFunctEq_8}
\end{equation}
Further, in Eq. \ref{eq:QuantumFieldTheoryInteractSchrodingerFieldForSevDiffKindOfIdPartSIndGreensFunctVertexFunctEq_8}
\begin{eqnarray} 
\frac{\delta{G_{k}\left(4,5\right)}}{\delta{V_{tot}\left(3,k\right)}} &=& \int d6 \frac{\delta{G_{k}\left(4,5\right)}}{\delta{\varphi\left(6\right)}} \frac{\delta{\varphi\left(6\right)}}{\delta{V_{tot}\left(3,k\right)}} \nonumber \\
&=& - \int d\bar{4} \int d\bar{5} \int d\bar{6}  G_{k}\left(4,\bar{4}\right)  \frac{\delta{G^{-1}_{k}\left(\bar{4},\bar{5}\right)}}{\delta{V_{tot}\left(\bar{6},k\right)}} \nonumber \\
&&\times \delta\left(\bar{6}-3\right) G_{k}\left(\bar{5},5\right) \nonumber \\
&=& - \int d\bar{4} \int d\bar{5} G_{k}\left(4,\bar{4}\right)  \frac{\delta{G^{-1}_{k}\left(\bar{4},\bar{5}\right)}}{\delta{V_{tot}\left(3,k\right)}} G_{k}\left(\bar{5},5\right) \nonumber \\
&=& \int d\bar{4} \int d\bar{5} G_{k}\left(4,\bar{4}\right) G_{k}\left(\bar{5},5\right) \Gamma_{k}\left(\bar{4},\bar{5},3\right),
\label{eq:QuantumFieldTheoryInteractSchrodingerFieldForSevDiffKindOfIdPartSIndGreensFunctVertexFunctEq_9}
\end{eqnarray}
where the following relation was used
\begin{equation} 
\int d6 \frac{\delta{V_{tot}\left(\bar{6},k\right)}}{\delta{\varphi\left(6\right)}}  \frac{\delta{\varphi\left(6\right)}}{\delta{V_{tot}\left(3,k\right)}} = \delta\left(\bar{6}-3\right).
\label{eq:QuantumFieldTheoryInteractSchrodingerFieldForSevDiffKindOfIdPartSIndGreensFunctVertexFunctEq_10}
\end{equation}
By using Eq. \ref{eq:QuantumFieldTheoryInteractSchrodingerFieldForSevDiffKindOfIdPartSIndGreensFunctVertexFunctEq_9} in Eq. \ref{eq:QuantumFieldTheoryInteractSchrodingerFieldForSevDiffKindOfIdPartSIndGreensFunctVertexFunctEq_8} (some relabeling)
\begin{eqnarray} 
\frac{\delta{\Sigma_{k}\left(1,2\right)}}{\delta{V_{tot}\left(3,k\right)}} &=& \int d4\int d5 \int d6 \int d7 \frac{\delta{\Sigma_{k}\left(1,2\right)}}{\delta{G_{k}\left(4,5\right)}} \nonumber \\
&&\times G_{k}\left(4,6\right) G_{k}\left(7,5\right) \Gamma_{k}\left(6,7,3\right).
\label{eq:QuantumFieldTheoryInteractSchrodingerFieldForSevDiffKindOfIdPartSIndGreensFunctVertexFunctEq_12}
\end{eqnarray}
Further, by using Eq. \ref{eq:QuantumFieldTheoryInteractSchrodingerFieldForSevDiffKindOfIdPartSIndGreensFunctVertexFunctEq_12} in Eq. \ref{eq:QuantumFieldTheoryInteractSchrodingerFieldForSevDiffKindOfIdPartSIndGreensFunctVertexFunctEq_7}
\begin{eqnarray} 
\Gamma_{k}\left(1,2,3\right) &=& \delta\left(1 - 2\right) \delta\left(1 - 3\right) \nonumber \\
&&+ \int d4\int d5 \int d6 \int d7 \frac{\delta{\Sigma_{k}\left(1,2\right)}}{\delta{G_{k}\left(4,5\right)}} \nonumber \\
&&\times G_{k}\left(4,6\right) G_{k}\left(7,5\right) \Gamma_{k}\left(6,7,3\right).
\label{eq:QuantumFieldTheoryInteractSchrodingerFieldForSevDiffKindOfIdPartSIndGreensFunctVertexFunctEq_13}
\end{eqnarray}
This is one of the Hedin's equations for a species $k$ and it is similar to the one given in Eq. (75) of Ref. \cite{Harkonen-ManybodyGreensFunctionTheoryOfElectronsAndNucleiBeyondTheBornOppenheimerApproximation-PhysRevB.101.235153-2020} for electrons only.

\subsection{Screened Coulomb Interaction}
\label{ScreenedCoulombInteraction}

By using Eqs. \ref{eq:EquationsOfMotionEq_17} and \ref{eq:EquationsOfMotionEq_22} [here $n_{k}\left(1\right) \equiv \left\langle  \hat{n}_{k}\left(1\right) \right\rangle$]
\begin{eqnarray} 
W_{k}\left(1,2\right) &=& Z^{2}_{k} \int d3 v\left(1,3\right) \left[\delta\left(2-3\right) + \int d4  v\left(2,4\right) \frac{\delta{\left\langle  \hat{n}\left(4\right) \right\rangle}}{\delta{\varphi\left(3\right)}}\right] \nonumber \\
&=& Z^{2}_{k} v\left(1,2\right) + Z^{2}_{k} \int d3  \int d4 v\left(1,3\right) \sum_{k'} Z_{k'} \frac{\delta{n_{k'}\left(4\right) }}{\delta{\varphi\left(3\right)}} v\left(2,4\right) \nonumber \\
&=& Z^{2}_{k} v\left(1,2\right) + Z^{2}_{k} \int d3  \int d4 \int d5 \sum_{k'} Z_{k'} v\left(1,3\right) \nonumber \\
&&\times \frac{\delta{n_{k'}\left(4\right)}}{\delta{V_{tot}\left(5,k'\right)}} \frac{\delta{V_{tot}\left(5,k'\right)}}{\delta{\varphi\left(3\right)}} v\left(2,4\right) \nonumber \\
&=& Z^{2}_{k} v\left(1,2\right) + Z^{2}_{k} \int d3  \int d4 \int d5 \sum_{k'} Z_{k'}  \nonumber \\
&&\times v\left(1,3\right) \frac{\delta{n_{k'}\left(4\right)}}{\delta{V_{tot}\left(5,k'\right)}} \epsilon^{-1}_{k'}\left(5,3\right) v\left(2,4\right) \nonumber \\
&=& Z^{2}_{k} v\left(1,2\right) \nonumber \\
&&+  Z^{2}_{k} \int d3 \int d4 v\left(2,3\right) \sum_{k'} \frac{\delta{n_{k'}\left(3\right)}}{\delta{V_{tot}\left(4,k'\right)}}  W_{k'}\left(1,4\right).
\label{eq:QuantumFieldTheoryInteractSchrodingerFieldForSevDiffKindOfIdPartSIndGreensFunctScrCoulIntEq_1}
\end{eqnarray}
Define
\begin{equation} 
P_{k'}\left(3,4\right) \equiv \frac{\delta{n_{k'}\left(3\right)}}{\delta{V_{tot}\left(4,k'\right)}},
\label{eq:QuantumFieldTheoryInteractSchrodingerFieldForSevDiffKindOfIdPartSIndGreensFunctScrCoulIntEq_2}
\end{equation}
and thus
\begin{eqnarray} 
W_{k}\left(1,2\right) &=& Z^{2}_{k} v\left(1,2\right) \nonumber \\
&&+  Z^{2}_{k} \int d3 \int d4  v\left(2,3\right) \sum_{k'} P_{k'}\left(3,4\right)  W_{k'}\left(1,4\right).
\label{eq:QuantumFieldTheoryInteractSchrodingerFieldForSevDiffKindOfIdPartSIndGreensFunctScrCoulIntEq_3}
\end{eqnarray}
The quantity $P_{k}\left(3,4\right)$ is sometimes called the polarization or polarization propagator (in this case for a species $k$). Note that only the charge $Z^{2}_{k}$ labels the screened interaction $W_{k}\left(1,2\right)$ for the species $k$.

\subsection{Polarizations}
\label{Polarizations}

Consider the polarization defined in Eq. \ref{eq:QuantumFieldTheoryInteractSchrodingerFieldForSevDiffKindOfIdPartSIndGreensFunctScrCoulIntEq_2}
\begin{equation} 
P_{k}\left(1,2\right) = \frac{\delta{ \left\langle  \hat{n}_{k}\left(1\right) \right\rangle }}{\delta{V_{tot}\left(2,k\right)}}.
\label{eq:QuantumFieldTheoryInteractSchrodingerFieldForSevDiffKindOfIdPartSIndGreensFunctPolarizEq_1}
\end{equation}
The electron density can be written in terms of one-body Green's function as
\begin{equation} 
\left\langle  \hat{n}_{k}\left(1\right) \right\rangle = -i \hbar G_{k}\left(1,1^{+}\right).
\label{eq:QuantumFieldTheoryInteractSchrodingerFieldForSevDiffKindOfIdPartSIndGreensFunctPolarizEq_2}
\end{equation}
Now by using Eq. \ref{eq:QuantumFieldTheoryInteractSchrodingerFieldForSevDiffKindOfIdPartSIndGreensFunctPolarizEq_2} in Eq. \ref{eq:QuantumFieldTheoryInteractSchrodingerFieldForSevDiffKindOfIdPartSIndGreensFunctPolarizEq_1}
\begin{equation} 
P_{k}\left(1,2\right) = -i \hbar \frac{\delta{  G_{k}\left(1,1^{+}\right) }}{\delta{V_{tot}\left(2,k\right)}}.
\label{eq:QuantumFieldTheoryInteractSchrodingerFieldForSevDiffKindOfIdPartSIndGreensFunctPolarizEq_3}
\end{equation}
Then, by using the result
\begin{equation} 
\frac{\delta{G_{k}\left(1,1^{+}\right)}}{\delta{V_{tot}\left(2,k\right)}} = - \int d4 \int d5 G\left(1,4\right) \frac{\delta{G^{-1}_{k}\left(4,5\right)}}{\delta{V_{tot}\left(2,k\right)}}  G\left(5,1^{+}\right),
\label{eq:QuantumFieldTheoryInteractSchrodingerFieldForSevDiffKindOfIdPartSIndGreensFunctPolarizEq_3_2}
\end{equation}
we can write Eq. \ref{eq:QuantumFieldTheoryInteractSchrodingerFieldForSevDiffKindOfIdPartSIndGreensFunctPolarizEq_3} as
\begin{equation} 
P_{k}\left(1,2\right) = i \hbar \sum_{\sigma_{1}} \int d4 \int d5  G_{k}\left(1,4\right) \frac{\delta{G^{-1}_{k}\left(4,5\right)}}{\delta{V_{tot}\left(2,k\right)}}  G_{k}\left(5,1^{+}\right),
\label{eq:QuantumFieldTheoryInteractSchrodingerFieldForSevDiffKindOfIdPartSIndGreensFunctPolarizEq_4}
\end{equation}
and by using the definition of the vertex function given by Eq. \ref{eq:EquationsOfMotionEq_19} (some relabeling done at the end)
\begin{equation} 
P_{k}\left(1,2\right) = -i \hbar \int d3 \int d4  G_{k}\left(1,3\right) G_{k}\left(4,1^{+}\right) \Gamma_{k}\left(3,4,2\right).
\label{eq:QuantumFieldTheoryInteractSchrodingerFieldForSevDiffKindOfIdPartSIndGreensFunctPolarizEq_6}
\end{equation}
Equation \ref{eq:QuantumFieldTheoryInteractSchrodingerFieldForSevDiffKindOfIdPartSIndGreensFunctPolarizEq_6} is similar to the one given in Eq. (75) of Ref. \cite{Harkonen-ManybodyGreensFunctionTheoryOfElectronsAndNucleiBeyondTheBornOppenheimerApproximation-PhysRevB.101.235153-2020} for electrons only. It is one of the Hedin's equations.


\section*{ORCID iDs}

Ville H\"{a}rk\"{o}nen: https://orcid.org/0000-0002-2956-5457

\section*{References}
\bibliography{bibfile}
\end{document}